\documentclass[aps,prd,twocolumn,amsmath,amssymb]{revtex4-2}
\usepackage{epsfig,xcolor,amsmath,tikz}
\usepackage{graphicx}
\usepackage{epsfig}
\usepackage{color}
\usepackage{physics}

\usepackage[colorlinks]{hyperref}
\usepackage{braket}
\hypersetup{linkcolor=blue,citecolor=blue,urlcolor=blue}

\begin{document}


\title{Cosmological Singularities and Quantum Particles } 


\author{Samuel W. P. Oliveira}
\email[]{sw.oliveira55@gmail.com}
\affiliation{PPGCosmo, Universidade Federal do Esp\'{\i}rito Santo,
	Vit\'oria, 29075-910, ES, Brazil} 
\author{Alexander Yu. Kamenshchik}
\email[]{kamenshchik@bo.infn.it}
\affiliation{Department of Physics and Astronomy ``A. Righi'', University of Bologna, via Irnerio 46,  40126 Bologna, Italy\\
INFN, section of Bologna, viale Berti Pichat 6/2, 40127 Bologna, Italy}

\date{\today}

\begin{abstract}
We study if there is an opportunity to describe quantum particles in the vicinity of three types of cosmological singularities,
big bang--big crunch, big rip and big brake. Writing down the Dirac equation for spinors, and choosing a convenient parametrization for basis functions of the spinor field, we show that the corresponding second-order differential equation has two independent solutions which are non-singular in the case of all three types of singularities. That permits us to construct the Fock space for the spinor particles and to interprete this fact as their opportunity to cross these cosmological singularities. We show also that this is impossible to do for scalar particles and changing the parametrization does not help. Thus, fermions look more resilient 
to the passage of the cosmological singularities than bosons. 
\end{abstract}



\hypersetup{
    urlcolor=black   
}
\maketitle
\hypersetup{urlcolor=blue}

\section{Introduction}
The problem of cosmological singularities has been attracting the attention of theoreticians
working in gravity and cosmology since the early 1950s \cite{Lif-Khal}. In the 1960s  general theorems
about the conditions for the appearance of singularities were proven \cite{Pen-Hawk,Pen} and the oscillatory
regime of approaching the singularity \cite{BKL}, called also ``Mixmaster universe'' \cite{Misner}, was discovered.
Until the end of 1990s almost all discussions about singularities were devoted to the big bang and big crunch singularities, which are characterized by a vanishing volume of the universe.
However, kinematical investigations of Friedmann cosmologies have raised the possibility
of sudden future singularity occurrence \cite{sudden}, characterized by a diverging  acceleration $\ddot{a}$, whereas both the
scale factor $a$ and  $\dot{a}$  are finite. Then, the Hubble parameter $H = \frac{\dot{a}}{a}$  and the energy density $\rho$
are also finite, while the first derivative of the Hubble parameter and the pressure $p$ diverge.
Interestingly, in some of the models of dark energy  sudden (soft)  future singularities arise quite naturally.
Thus, during last years a plenty of work, devoted to the study of future singularities was done (see, for example, the review \cite{my-review} and references therein). 
A distinguishing feature of these singularities is the fact that rather often they can be crossed \cite{Fern,Fern1}. Indeed, due to the finiteness of the Christoffel symbols, the geodesic equations are not singular and, hence, can be continued through these singularities without difficulties.   

On the other hand, the crossing of the big bang -- big crunch singularity is more complicated and looks more counter-intuitive with respect to the crossing of the soft future singularities. However, during last years some approaches to this problem were elaborated \cite{Bars,Bars1,Wetterich,Wetterich1,Prester,Prester1,we-cross,we-cross1,we-cross2,we-cross3}. Behind of these approaches there are basically two general ideas. Firstly, to cross the singularity means to give a prescription matching non-singular, finite quantities before and after such a crossing. Secondly, such a description can be achieved by using a convenient choice of the field parametrization. 

In connection with the investigations of the nature of the cosmological singularities and of the opportunities of their crossing, it is interesting to study what happens with the quantum particles in the vicinity of a singularity and in the process of their crossing. The question what happens with particles (in quantum field theoretical sense) when the universes passes through the cosmological singularity was studied in \cite{Olesya},
where all considerations were carried for a  scalar field in a flat Friedmann universe satisfying the Klein-Gordon equation:
\begin{equation}
\Box\phi +V'(\phi) = 0.
\label{K-G}
\end{equation} 
One can consider a spatially homogeneous solution of this equation $\phi_0$, depending only on time $t$ as a classical background.
A small deviation from this background solution can be represented as a sum of  Fourier harmonics satisfying linearised equations
\begin{equation}
\ddot{\phi}(\vec{k},t)+3\frac{\dot{a}}{a}\dot{\phi}(\vec{k},t)+\frac{\vec{k}^2}{a^2}\phi(\vec{k},t) + V''(\phi_0(t))\phi(\vec{k},t) = 0.
\label{K-G1}
\end{equation}
The corresponding quantised field is 
\begin{equation}
\hat{\phi}(\vec{x},t) = \int d^3\vec{k}\Big(\hat{a}(\vec{k})u(k,t)e^{i\vec{k}\cdot\vec{x}}+\hat{a}^+(\vec{k})u^*(k,t)e^{-i\vec{k}\cdot\vec{x}}\Big),
\label{quant}
\end{equation}  
where the creation and the annihilation operators satisfy the standard commutation relations:
\begin{equation}
[\hat{a}(\vec{k}), \hat{a}^+(\vec{k}')]=\delta(\vec{k}-\vec{k}').
\label{quant1}
\end{equation}
The basis functions should be normalised 
\begin{equation}
u(k,t)\dot{u}^*(k,t)-u^*(k,t)\dot{u}(k,t)=\frac{i}{(2\pi)^3a^3(t)},
\label{canon2}
\end{equation}
so that the canonical commutation relations between the field $\phi$ and its canonically conjugate momentum 
 $\hat{{\cal P}}$ 
\begin{equation}
[\hat{\phi}(\vec{x},t), \hat{{\cal P}}(\vec{y},t')]=i\delta(\vec{x}-\vec{y})
\label{canon}
\end{equation}
were satisfied.

The linearised Klein-Gordon equation has two independent solutions, which must be non-singular to define a particle.
 It is a non-trivial requirement in the situations when a singularity or other kind of irregularity of the spacetime geometry occurs.
 The existence of this set of solutions and, hence, an opportunity to construct the Fock vacuum for quantum scalar particles through
 different kinds of cosmological singularities
 was studied in \cite{Olesya}.
 All the calculations were done in a certain fixed parametrization 
 of the scalar field. At the same time it is clear that redefining the basis functions, by multiplying them by some power of the scale factor, one redefines 
 the very notion of particle and, in this case, the behavior of the new defined particles in the vicinity of the singularities can also be changed. Note that the consequences of such a kind of reparametrization were studied earlier  in the context of the decoherence effects in quantum cosmology \cite{decoh,decoh1}.   
 In the present paper we shall study how fermion particles behave in the vicinity of different types of cosmological singularities. Besides, we come back to the case 
 of the scalar particles studied in \cite{Olesya}, to see what happens in the case when the parametrization of the harmonics of the scalar field is changed. 
 Generally, we shall show that the fermions are more resilient to different kinds of singularities than bosons. The second section of the paper is devoted to the fermions,
 the third one -- to scalars, while the fourth section contains some concluding remarks.
 
 \section{Fermions in cosmology and singularities}
 
 The behavior of fermions in quantum cosmology or more generally, in curved spacetimes was considered in many papers (see e.g. \cite{ferm-cosm,ferm-cosm1,ferm-cosm2,Kam-Mish,Kam-Mish1,Samuel}). Let us write down some basic formula which we shall use. 
 The action for the Dirac fermions in a curved spacetime is
\begin{equation}
S_f = \int d^4x \sqrt{-g} \Bigg[ \frac{i}{2} \Big( \bar{\psi}\gamma^\mu\nabla_\mu\psi - \nabla_\mu\bar{\psi}\gamma^\mu\psi
\Big) - m\bar{\psi}\psi \Bigg].
\label{action}
\end{equation}
We shall consider a flat Friedmann universe with the metric 
\begin{equation}
ds^2=dt^2-a^2(t)d\vec{x}^{\,2}.
\label{Fried}
\end{equation}
The Dirac equation following from the action \eqref{action} with the metric \eqref{Fried} is
\begin{equation}
\Bigg[i\gamma^0 \bigg( \partial_t+\frac{3}{2}\frac{\dot a}{a} \bigg) + \frac{i}{a}\gamma^i\partial_i - m \Bigg]\psi=0.
\label{Dirac}
\end{equation}
The spinor field operator $\psi$ can be represented in the following form:
\begin{widetext}
\begin{equation}
\hat{\psi}(\vec{x},t) = a^{-3/2}(t)
\int \frac{d^3k}{(2\pi)^{3/2}}
\sum_{s=1}^{2} \bigg[ \hat{a}_s(\vec{k})u_s(\vec{k}, t)e^{i\vec{k}\cdot\vec{x}} + \hat{b}_s^\dagger(\vec{k})v_s(\vec{k}, t)e^{-i\vec{k}\cdot\vec{x}} \bigg].
\label{Dirac1}
\end{equation}
\end{widetext}
Here $\hat{a}_s(\vec{k}), \hat{a}_s^\dagger(\vec{k})$ and $\hat{b}_s(\vec{k}), \hat{b}_s^\dagger(\vec{k})$ are the operators of annihilation and creation for particles and antiparticles respectfully, while $u_s(\vec{k}, t)$ and $v_s(\vec{k}, t)$ are the corresponding basis functions. 
The weight factor $a^{-3/2}$  is especially chosen in order to write the action \eqref{action}
in a special parametrization reflecting the conformal properties of the spinor field. It is
conformally invariant in the massless case and, therefore, decouples in this parametrization
from the gravitational background.

Each four-component Dirac spinor consists of two two-component Weyl spinors:
\begin{equation}
u_s(\vec k,t)
=
\begin{pmatrix}
u_k(t)\xi_s(\vec k)\\
-\tilde u_k(t)\xi_s(\vec k)
\end{pmatrix},
\qquad
v_s(\vec k,t)
=
\begin{pmatrix}
v_k(t)\xi_s(\vec k)\\
-\tilde v_k(t)\xi_s(\vec k)
\end{pmatrix},
\end{equation}
where the two-component spinors $\xi(\vec{k})$ satisfy the following equations  
\begin{equation}
\vec{\sigma}\cdot\hat{k}\,\xi_s(\vec{k})=\lambda_s\,\xi_s(\vec{k}),\qquad \lambda_s=\pm1,
\end{equation}
and to  the normalization condition
\begin{equation}
\xi_s^\dagger(\vec{k})\xi_{s'}(\vec{k})=\delta_{ss'}.
\end{equation}

The basis functions in Eq.~\eqref{Dirac1} satisfy the following equations:
\begin{equation}
i\partial_t\,u_s(\vec{k}, t) = \Bigg( \frac{\vec{\alpha}\cdot\vec{k}}{a(t)} + \beta m\Bigg)  u_s(\vec{k}, t),
\label{Dirac2}
\end{equation}
\begin{equation}
i\partial_t\,v_s(\vec{k}, t) = \Bigg(- \frac{\vec{\alpha}\cdot\vec{k}}{a(t)} + \beta m\Bigg)  v_s(\vec{k}, t).
\label{Dirac3}
\end{equation}
In the Weyl representation 
\begin{equation}
    \beta=\left(
    \begin{tabular}{cc}
        0 & I \\
        I & 0
    \end{tabular}
    \right),
    \qquad
    \vec{\alpha}=\left(
    \begin{tabular}{cc}
         $-\vec{\sigma}$ & 0 \\
         0 & $\vec{\sigma}$
    \end{tabular}
    \right),
\label{Weyl}
\end{equation}
Eqs.~\eqref{Dirac2} and \eqref{Dirac3} imply the following equations for
the four scalar functions: 
\begin{equation}
\begin{aligned}
i\dot{u}_k &= -\nu_{k} u_k-m\tilde{u}_k,\\
i\dot{\tilde{u}}_k &= \nu_{k}\tilde{u}_k-mu_k,
\end{aligned}
\label{dif-eq}
\end{equation}
\begin{equation}
\begin{aligned}
i\dot{v}_k &= \nu_{k} v_k-m\tilde{v}_k,\\
i\dot{\tilde{v}}_k &= -\nu_{k}\tilde{v}_k-mv_k,
\end{aligned}
\label{dif-eq1}
\end{equation}
where 
\begin{equation}
\nu_{k}(t) = \frac{k}{a(t)}.
\label{nu}
\end{equation}

The corresponding second order equation for the scalar function $u_k$ is 
\begin{equation}
\ddot{u}_k+ \big( -i\dot{\nu_{k}} +\nu_{k}^2 +m^2 \big)u_k=0,
\label{dif-eq2}
\end{equation}
and for the scalar function $v_k$
\begin{equation}
\ddot{v}_k+ \big( i\dot{\nu_{k}} +\nu_{k}^2 +m^2 \big)v_k=0.
\label{dif-eq3}
\end{equation}

Now we are in a position to study the behavior of fermion quantum particles in the vicinity of different kinds of singularities.
It will be convenient to rewrite Eqs.~\eqref{dif-eq2} and \eqref{dif-eq3} in a more explicit form
 \begin{equation}
\ddot{u}_k+ \left( i\frac{k\dot{a}}{a^2} +\frac{k^2}{a^2} +m^2 \right)u_k=0,
\label{dif-eq4}
\end{equation}
\begin{equation}
\ddot{v}_k+ \left( -i\frac{k\dot{a}}{a^2} +\frac{k^2}{a^2} +m^2 \right)v_k=0.
\label{dif-eq5}
\end{equation}
It is enough to consider one of these two equations, because the asymptotic behavior of their solutions in the vicinity of singularity is the same. 

\subsection{Big bang -- big crunch singularity}

Let us consider what happens with fermion particles in the vicinity (or at the crossing) of the big bang -- big crunch singularity.
In the universe filled with a perfect fluid with the constant equation of state parameter
\begin{equation}
p = w\rho, \quad -\frac13 < w \leq 1,
\label{state}
\end{equation}
where as usual, $p$ is the pressure and $\rho$ is the energy density, the scale factor behaves as
\begin{equation}
a(t) = a_0t^{\frac{2}{3(1+w)}}.
\label{Fried1}
\end{equation}
Substituting  expression \eqref{Fried1} into Eq. \eqref{dif-eq4}, we obtain
\begin{equation}
    \ddot{u}_k+\bigg[\frac{2ik}{3(1+w)a_0}t^{-\frac{5+3w}{3(1+w)}}+\frac{k^2}{a_0^2}t^{-\frac{4}{3(1+w)}}+m^2\bigg]u_k=0.
\label{dif-eq6}    
\end{equation}
In the vicinity of the big bang or big crunch singularity, where $t \rightarrow 0$, the first term in the square brackets diverges strongly than the second one and Eq. \eqref{dif-eq6} reduces to 
\begin{equation}
    \ddot{u}_k+\frac{2ik}{3(1+w)a_0}t^{-\frac{5+3w}{3(1+w)}}u_k=0.
\label{dif-eq7}    
\end{equation}
Its solution can be expressed in terms of Bessel functions 
\begin{equation}
    u_k(t)=C_1\sqrt{t}\,J_\zeta(\xi)+C_2\sqrt{t}\,Y_\zeta(\xi),
\label{Bessel}
\end{equation}
where
\begin{eqnarray}
 &&  \zeta={\frac{3(1+w)}{1+3w}}, \qquad \xi(t)=\frac{6(1+w)\sqrt{\lambda}}{1+3w}\,t^{\frac{1+3w}{6(1+w)}},\nonumber \\ &&\lambda=\frac{2ik}{3(1+w)a_0}.
    \label{Bessel1}
\end{eqnarray}
It is easy to see that in the vicinity of the singularity, when $t \rightarrow 0$, one has 
\begin{equation}
\sqrt{t}\,J_\zeta(\xi) \sim t,
\label{Bessel2}
\end{equation}
\begin{equation}
\sqrt{t}\,Y_\zeta(\xi) \sim const,
\label{Bessel3}
\end{equation}
and, hence, having two independent non-singular solutions we can construct the Fock space for fermion particles crossing the big bang -- big crunch singularity.

\subsection{Big rip singularity}

The big rip singularity was discovered during the analysis of different dark energy models \cite{Rip,Rip1}.
The simplest situation when such a singularity appears is the Friedmann universe filled with the so called phantom 
fluid, which the equation of state parameter in Eq.~\eqref{state} is less than $-1$, i.e. $w < -1$. In this case the expansion of the universe arrives, during a finite interval of the cosmic time, to the situation when the scale factor, its time derivative and the scalar curvature all tend to infinity.   It is convenient to represent the cosmological evolution in this case as follows
\begin{equation}
    a(t)=a_0\,(-t)^{\frac{2}{3(1+w)}}, \qquad w<-1,
    \label{a.b.rip}
\end{equation}
where the time parameter $t$ is negative and when it tends to zero, the universe encounters the big rip singularity. 
We can restrict ourselves by the case when the equation of state parameter is not far away from $-1$, namely
\begin{equation}
-\frac53 < w < -1.
\label{phantom}
\end{equation}
The structure of the second-order differential equation for the function $u_{k}$ coincides with that of Eq. \eqref{dif-eq6}.
However, in this case both the time-dependent terms in the square brackets tend to zero, while the mass term survives, reducing Eq. \eqref{dif-eq6} to 
\begin{equation}
\ddot{u}_{k} + m^2u_{k}=0,
\label{phantom1}
\end{equation}
with the solution
\begin{equation}
u_k(t) = C_1 \sin mt + C_2\cos mt.
\label{phantom2}
\end{equation}
Thus, in this case we have again two non-singular independent solutions of our differential equation in the vicinity of singularity and can construct the Fock space.

\subsection{Big brake singularity}

The big brake singularity belongs to the so called soft or sudden future singularities. It was firstly described in paper \cite{Brake} and its study developed in papers \cite{Brake1, Brake2, Brake3}. When the universe encounters this singularity, its scale factor 
acquires some finite value, the time derivative is equal to zero, while the second time derivative of the scale factor tends to minus infinity. Thus, the universe's expansion has an infinite deceleration. The scalar curvature at this moment is infinite, which the Christoffel symbols are equal to zero. A rather simple case of the enecounter of the universe with the big brake singularity  can be seen if its is filled with the anti-Chaplygin gas \cite{Brake} having the equation of state 
\begin{equation}
p = \frac{A}{\rho}.
\label{anti-Chap}
\end{equation}
Note that in Eq. \eqref{anti-Chap}, the constant $A$ and the pressure are both positive, which distinguishes this model from that with the Chaplygin gas with the equation of state $p~=~-\frac{A}{\rho}$ (see \cite{Chap}). The equation of state \eqref{anti-Chap} 
arises in a natural way in the theory of wiggly strings \cite{wiggly, wiggly1}.

It is easy to show that the dependence of the energy density of the anti-Chaplygin gas on the scale factor is given by the formula
\begin{equation}
\rho = \sqrt{\frac{B}{a^6}-A},
\label{anti-Chap1}
\end{equation}
where $B$ is a positive constant  representing an initial condition. The scale factor has a maximal value 
\begin{equation}
a_0 = \left(\frac{B}{A}\right)^{\frac16},
\label{scale-anti}
\end{equation}
which it acquires at some finite moment of time $t_{BB}$. Approaching the big brake, the scale factor behaves as 
\begin{equation}
a(t) = a_0 - C(t-t_{BB})^{\frac43},
\label{anti-Chap2}
\end{equation}
where $C = 2^{-7/3}\,3^{5/3}\,(AB)^{1/6}$. Thus, in the vicinity of the big brake Eq. \eqref{dif-eq4} becomes 
\begin{equation}
\ddot{u}_k +\left(\frac{k^2}{a_0^2} + m^2\right)u_k = 0.
\label{anti-Chap3}
\end{equation}
There are again two independent non-singular solutions given by 
\begin{equation}
u_k(t) = C_1\cos\sqrt{\frac{k^2}{a_0^2} + m^2}\ \ t+C_2\sin\sqrt{\frac{k^2}{a_0^2} + m^2}\ \ t,
\label{anti-Chap4}
\end{equation}
and the fermion particles safely pass through big brake singularity.\\
\vspace{1cm}

\section{Scalar particles revisited}

We have seen in the preceding section that quantum fermion particles pass through all considered types of cosmological singularities,
big bang -- big crunch, big rip and big brake. It is in contrast with result obtained in the preceding paper \cite{Olesya}, where it was shown that for the scalar particles it was impossible to cross big bang -- big crunch singularity. What is the reason of this difference? Can it be connected with the fact that for fermions we have used a particular ``conformal parametrization'' of basis functions? To answer this question let us try to reconsider the behavior of scalar particles in the vicinity of singularities, considering a general family of parametrizations. A similar approach was applied in another context earlier \cite{decoh},
where it was shown that by choosing a reasonable parametrization of the scalar perturbations in quantum cosmology, one can construct the reduced density matrix for the quantum state of the universe, possessing good  decoherence properties. 

Instead of Eq. \eqref{quant}, consider the following representation for the scalar field on the cosmological background:
\begin{eqnarray}
\hat{\phi}(\vec{x},t) =a^{-\mu}(t) \int d^3\vec{k}\ \Big[\hat{a}(\vec{k})u(k,t)e^{i\vec{k}\cdot\vec{x}}
\nonumber \\
+\hat{a}^+(\vec{k})u^*(k,t)e^{-i\vec{k}\cdot\vec{x}}\Big].
\label{quant-new}
\end{eqnarray}  
In this case the second-order differential equation for the basis function $u_k$ looks as follows:
\begin{widetext}
\begin{equation}
    \ddot{u}_k
    +(3-2\mu)\frac{\dot{a}}{a}\,\dot{u}_k
    +\left[
    \mu(\mu-2)\left(\frac{\dot{a}}{a}\right)^2-\mu\frac{\ddot{a}}{a}
    +\frac{k^2}{a^2}
    +m_\phi^2
    \right]u_k=0.
    \label{v.general}
\end{equation}
\end{widetext}
Two special cases are particularly interesting. For ${\mu=3/2}$, the term with the first derivative  disappears. It also yields a scale-independent Wronskian ${W\propto a^0}$, recovering the normalization structure of Minkowski spacetime. The second case is for the conformal-weight rescaling ${\mu=1}$, which is the parametrization most naturally adapted to conformal symmetry in the massless limit. Now we can study the behavior of the solutions of Eq.~\eqref{v.general} in the vicinity of different types of singularities. 

\subsection{Big bang -- big crunch singularity}

In the vicinity of the big bang -- big crunch singularity Eq.~\eqref{v.general} looks as follows:
\begin{widetext}
\begin{equation}
    \ddot{u}_k
    +\frac{2(3-2\mu)}{3(1+w)}\,t^{-1}\,\dot{u}_k
    +\bigg[
    \frac{4\mu(\mu-2)+2\mu(3w+1)}{9(1+w)^2}\,t^{-2}
    +\frac{k^2}{a_0^2}\,t^{-\frac{4}{3(1+w)}}
    +m_\phi^2 \bigg]u_k=0.
\end{equation}
\end{widetext}
When $t \rightarrow 0$ we have
\begin{equation}
    \ddot{u}_k
    +\mathcal{A}\,t^{-1}\dot{u}_k +\mathcal{B}\,t^{-2}u_k=0,
    \label{v.bb.asym}
\end{equation}
where
\begin{eqnarray}
    &&\mathcal{A}=\frac{2(3-2\mu)}{3(1+w)},\nonumber \\
   && \mathcal{B}=\frac{4\mu(\mu-2)+2\mu(3w+1)}{9(1+w)^2}.
\end{eqnarray}
Hence, the solutions can be written explicitly as
\begin{equation}
    u_k(t)=c_1t^{\alpha_1}+c_2t^{\alpha_2}, 
    \label{sol0}
\end{equation}    
    where
    \begin{equation} 
    \alpha_1=\frac{2\mu}{3(1+w)},\quad 
   \alpha_2=\frac{2\mu+3(w-1)}{3(1+w)}.
    \label{v.b.bang.sol.}
\end{equation}
The regularity of the two independent solutions is  connected to the values of the  $\mu$ parameter. For ${\mu=3/2}$, one obtains
\begin{equation}
    u^{(5)}_k(t)\sim c_1\,t^{\frac{1}{1+w}}
    +c_2\,t^{\frac{w}{1+w}}.
\label{sol}
\end{equation}
If $w < 0$ then the second solution diverges at ${t \rightarrow 0}$. However, if $w > 0$, both solutions tend to zero at 
${t \rightarrow 0}$ and it is not good as well. Indeed, the cosmological wave function describing the corresponding $u_k$ mode has the following form (see e.g. \cite{Olesya} and the references therein):
\begin{equation}
\Psi(f_k) = \frac{1}{\sqrt{|u_k(t)|}}\exp\left(\frac{ia^3(t)\dot{u}_k^*(t)f_k^2}{2u_k^*(t)}\right),
\label{wave}
\end{equation}
where $f_k$ plays the role of the coordinate in the functional configuration space of harmonics of perturbations.
If we are not able to construct the basis function which tends to a constant, when $t \rightarrow 0$,  function \eqref{wave} does not behave well. Thus, the only case when solution \eqref{sol} can serve for the construction of the Fock space is the case 
of the universe filled with the dust-like matter with $w=0$. 

For the same reason, when we consider the case $\mu = 1$, the solution 
\begin{equation}
    u_k\sim c_1\,t^{\frac{2}{3(1+w)}}+c_2\,t^{\frac{3w-1}{3(1+w)}},
\label{sol1}
\end{equation}
can serve for the construction of the Fock basis for the scalar particles only for $w = \frac13$. \\
\vspace{1cm}
 
\subsection{Big rip singularity}

In the vicinity of the big rip singularity expressions \eqref{sol0} and \eqref{v.b.bang.sol.} are still valid, but now our equation of state parameter is $w < -1$. Thus, the exponent $\alpha_1$ is always negative for $\mu > 0$ and the corresponding solution $t^{\alpha_1}$ diverges.
Hence, the construction of the Fock space in the vicinity of the big rip singularity is impossible. Let us note that in our old ``naive'' parametrization ${\mu = 0}$ used in \cite{Olesya}, $\alpha_1 = 0,\ \alpha_2 > 0$ for $w < -1$ and, hence, particles can pass through the big rip singularity.\\

\subsection{Big brake singularity}

In the vicinity of the big brake singularity, Eq.~\eqref{v.general} acquires a very simple form 
\begin{eqnarray}
    &&u''_k+\lambda_\mu\,\tau^{-2/3}v_k=0,\nonumber \\
    &&\lambda_\mu=\frac{4\mu C}{9a_0},
\label{scal-brake}
\end{eqnarray}
where $\tau = t_{BB}-t$ and prime means the differentiation with respect to $\tau$. 
The solution of Eq.~\eqref{scal-brake} is 
\begin{eqnarray}
    u_k(\tau)&=&\,
    c_1\,\sqrt{\tau}\, J_{3/4}\!\left(\frac{3}{2}\sqrt{\lambda_\mu}\,\tau^{2/3}\right)
    \nonumber \\
    &&+\,c_2\,\sqrt{\tau}\, Y_{3/4}\!\left(\frac{3}{2}\sqrt{\lambda_\mu}\,\tau^{2/3}\right).
    \label{v.sol.b.brake}
\end{eqnarray}
One can see that at $\tau \rightarrow 0$, one of the solutions 
\begin{equation}
\sqrt{\tau}\,J_{3/4}\!\left(\frac{3}{2}\sqrt{\lambda_\mu}\,\tau^{2/3}\right) \sim \sqrt{\tau},
\end{equation}
while other one 
\begin{equation}
\sqrt{\tau}\,Y_{3/4}\!\left(\frac{3}{2}\sqrt{\lambda_\mu}\,\tau^{2/3}\right) \rightarrow const.
\end{equation}
Thus, in this case we again can construct a Fock space for the particles crossing the big brake singularity.\\

\section{Concluding remarks}
In this work, we have investigated the behavior of fermionic and scalar fields in the vicinity of different types of cosmological singularities. Our main focus was to analyze the asymptotic regularity of the corresponding solutions to the equations of motion, under a convenient parametrization, in order to define a particle. It is required two independent non-singular solutions, to construct the Fock vacuum.

We have shown that by choosing a ``natural'' or conformal parametrization for  spinor particles, we can describe their passage through three types of cosmological singularities, big bang -- big crunch, big rip and big brake, while using different parametrizations for scalar particles, we cannot do it. Thus, it looks like fermions are more resilient with respect to passage through cosmological singularities than bosons. It would be interesting to understand if there are some deep reasons behind this ``empirical'' fact.

 \vspace{1cm}
 \acknowledgements
S.W.P.O.
is grateful to Coordena\c{c}\~{a}o de Aperfei\c{c}oamento de Pessoal
de N\'{i}vel Superior—CAPES (Brazil) for supporting his current Ph.D. project, and to Funda\c{c}\~{a}o de Amparo \`{a} Pesquisa e Inova\c{c}\~{a}o do Esp\'{i}rito Santo—FAPES (Brazil) and Dipartimento di Fisica e Astronomia dell'Universit\`a di Bologna and INFN, Sezione di Bologna (Italy) for supporting this project.


\begin{thebibliography}{99}
\bibitem{Lif-Khal}
E. M. Lifshitz and I. M.  Khalatnikov,  Investigations in relativistic cosmology,  Adv. Phys. {\bf 12}, 185 (1963).
\bibitem{Pen-Hawk}
S. W. Hawking and R. Penrose,   The singularities of gravitational collapse and cosmology, Proc. R. Soc.
Lond. A {\bf 314}, 529 (1970).
\bibitem{Pen}
R. Penrose, Structure of Space-Time. Benjamin, New York (1970).
\bibitem{BKL}
V. A. Belinsky, I. M.  Khalatnikov and E. M.  Lifshitz,  Oscillatory approach to a singular point in the relativistic
cosmology, Adv. Phys. {\bf 19}, 525 (1970).
\bibitem{Misner}
C. W. Misner, Mixmaster universe, Phys. Rev. Lett. {\bf 22}, 1071 (1969).
\bibitem{sudden}
J. D. Barrow, G. J.  Galloway and F. J. Tipler,   The closed-universe recollapse conjecture. Mon. Not. R. Astron.
Soc. {\bf 223}, 835 (1986).
\bibitem{my-review}
A. Y. Kamenshchik,  Quantum cosmology and late-time singularities, Class. Quantum Grav. {\bf 30}, 173001 (2013).
\bibitem{Fern}
L. Fernandez-Jambrina and R. Lazkoz,  Geodesic behaviour of sudden future singularities, Phys. Rev.
D {\bf 70}, 121503 (2004).
\bibitem{Fern1}
L. Fernandez-Jambrina and R. Lazkoz,   Classification of cosmological milestones, Phys. Rev. D {\bf 74} 064030 (2006).
\bibitem{Bars}
I. Bars, S. H.  Chen, P. J.  Steinhardt and N. Turok,  Antigravity and the big crunch/big bang transition, Phys.
Lett. B {bf 715}, 278 (2012).
\bibitem{Bars1}
 I. Bars, P.  Steinhardt and N. Turok, 
  Sailing through the big crunch-big bang transition,
  Phys.\ Rev.\ D {\bf 89},  061302 (2014).
\bibitem{Wetterich}
C. Wetterich, 
 Variable gravity Universe,
  Phys.\ Rev.\ D {\bf 89},  024005  (2014).
  \bibitem{Wetterich1}
  C. Wetterich, Eternal Universe,
  Phys.\ Rev.\ D {\bf 90},  043520  (2014).
\bibitem{Prester}
P. Dominis Prester, 
Curing black hole singularities with local scale invariance,
  Adv.\ Math.\ Phys.\  {\bf 2016},  6095236 (2016).
\bibitem{Prester1}
P. Dominis Prester, 
 Field redefinitions, Weyl invariance and the nature of mavericks,
  Class.\ Quant.\ Grav.\  {\bf 31},  155006 (2014).
  \bibitem{we-cross}
A. Y. Kamenshchik, E.O. Pozdeeva, S. Y.  Vernov, A. Tronconi and  G. Venturi,
 Transformations between Jordan and Einstein frames: Bounces, antigravity, and crossing singularities,
  Phys.\ Rev.\ D {\bf 94},  063510 (2016). 
\bibitem{we-cross1}
A. Y. Kamenshchik, E. O.  Pozdeeva, S. Y. Vernov, A. Tronconi and G. Venturi, 
Bianchi-I cosmological model and crossing singularities,
  Phys.\ Rev.\ D {\bf 95},  083503 (2017).
  \bibitem{we-cross2}
A. Y. Kamenshchik, E. O.  Pozdeeva, A. A.  Starobinsky, A. Tronconi,  G. Venturi and  S. Y. Vernov,   
Induced gravity, and minimally and conformally coupled scalar fields in Bianchi-I cosmological models,
Phys. Rev. D {\bf 97} 2, 023536  (2018).
\bibitem{we-cross3}
A. Y. Kamenshchik, E. O.  Pozdeeva, A. A.  Starobinsky, A. Tronconi, T. Vardanyan, G.  Venturi and S.Y.  Vernov, 
Duality between static spherically or hyperbolically symmetric solutions and cosmological solutions in scalar-tensor gravity,
Phys. Rev. D {\bf 98} 12, 124028  (2018).
\bibitem{Olesya}
O. Galkina and  A. Yu. Kamenshchik,  Future soft singularities, Born-Infeld-like fields and particles,     Phys. Rev. D {\bf 102} 2, 024078  (2020). 
\bibitem{decoh}
A. O. Barvinsky, A. Yu. Kamenshchik, C. Kiefer and I. V. Mishakov, Decoherence in quantum cosmology at the onset of inflation, Nucl. Phys. B {\bf 551}, 374  (1999).
\bibitem{decoh1}
A. O. Barvinsky, A. Yu. Kamenshchik and  C. Kiefer, Effective action and decoherence by fermions in quantum cosmology, Nucl. Phys. B {\bf 552}, 420  (1999). 
\bibitem{ferm-cosm}
P. D. D'Eath and J. J. Halliwell, Fermions in Quantum Cosmology,  Phys. Rev. D {\bf 35}, 1100  (1987). 
\bibitem{ferm-cosm1}
P. D. D'Eath and G. Esposito, Local boundary conditions for the Dirac operator and one loop quantum cosmology, Phys. Rev. D {\bf 43}, 3234  (1991). 
\bibitem{ferm-cosm2}
P. D. D'Eath and G. Esposito, Spectral boundary conditions in one loop quantum cosmology, Phys. Rev. D {\bf 44}, 1713  (1991).
\bibitem{Kam-Mish}
A. Yu. Kamenshchik and I. V. Mishakov, Fermions in one loop quantum cosmology,     Phys. Rev. D {\bf 47}, 1380  (1993). 
\bibitem{Kam-Mish1}
A. Yu. Kamenshchik and I. V. Mishakov, Fermions in one loop quantum cosmology 2: The Problem of correspondence between covariant and noncovariant formalisms, Phys. Rev. D {\bf 49}, 816  (1994).
\bibitem{Samuel}
S. W. P. Oliveira, G. Y. Oyadomari and I. L. Shapiro, Pauli equation and charged spin-1/2 particle in a weak gravitational field, Phys. Rev. D {\bf 110}, 105005 (2024). 
 \bibitem{Rip}
R. R. Caldwell, A Phantom menace?, Phys. Lett. B {\bf 545}, 23  (2002). 
\bibitem{Rip1}
R. R. Caldwell, M. Kamionkowski and N. N. Weinberg, Phantom energy and cosmic doomsday,  Phys. Rev. Lett. {\bf 91}, 071301  (2003).
\bibitem{Brake}
V. Gorini, A. Yu. Kamenshchik, U. Moschella and V. Pasquier, Tachyons, scalar fields and cosmology,     Phys. Rev. D {\bf 69},  123512 (2004). 
\bibitem{Brake1}
Z. Keresztes, L. A. Gergely, V. Gorini, U. Moschella and A. Yu. Kamenshchik, Tachyon cosmology, supernovae data and the Big Brake singularity,     Phys. Rev. D {\bf 79}, 083504  (2009).
\bibitem{Brake2}
Z. Keresztes, L. A. Gergely, A. Yu. Kamenshchik, V. Gorini and D. Polarski, Will the tachyonic Universe survive the Big Brake?,
Phys. Rev. D {\bf 82}, 123534  (2010).
\bibitem{Brake3}
Z. Keresztes, L. A. Gergely, A. Yu. Kamenshchik, V. Gorini and D. Polarski, Soft singularity crossing and transformation of matter properties,     Phys. Rev. D {\bf 88}, 023535  (2013). 
\bibitem{Chap}
A. Yu. Kamenshchik, U. Moschella and V. Pasquier, An alternative to quintessence, Phys. Lett. B {\bf 511}, 265 (2001). 
\bibitem{wiggly}
B. Carter, Duality Relation Between Charged Elastic Strings and Superconducting Cosmic Strings, Phys. Lett. B {\bf 224}, 61  (1989).
\bibitem{wiggly1}
A. Vilenkin, Effect of Small Scale Structure on the Dynamics of Cosmic Strings, Phys. Rev. D {\bf 41}, 3038  (1990). 
\end{thebibliography}
\end{document}